\documentclass[final,1p,times]{elsarticle}

\usepackage{amssymb}

\begin{document}

\begin{frontmatter}



\title{Global wake instabilities of low aspect-ratio flate-plates}


\author{O. Marquet \& M. Larsson}
\address{ONERA, The French Aerospace Lab \\
Experimental and Fundamental Aerodnyamics Department,\\ 
8 rue des Vertugadins, 92190 Meudon, France}

\begin{abstract}
This paper investigates the linear destabilization of three-dimensional 
steady wakes developing behind flate plates placed normal to the incoming 
flow. Plates characterized by low length-to-width ratio $L$ are considered here. 
By varying this aspect ratio in the range $1 \le L \le 6$ three destabilization scenarios are identified. 
For very low aspect ratio $1 \le L \le 2$, the flow is first destabilized, when increasing the Reynolds number,
by a steady global mode that breaks the top/bottom planar reflectional symmetry. The symmetric steady flow 
bifurcates, via a pitchfork bifurcation, towards an asymmetric steady wakes, similarly to the case of axisymmetric wakes behind sphere and disks. For long aspect ratio, $2.5 \le L \le 6$, the first unstable mode also breaks the top/bottom symmetry but is unsteady. A Hopf bifurcation occurs, as for the wake developing behind a two-dimensional circular cylinder. 
Finally an intermediate regime $2 \le L \le 2.5$ is found for which the flow gets first unstable to 
an unsteady mode that breaks the left/right planar reflectional symmetry.
\end{abstract}


\end{frontmatter}


\section{Introduction}

Wake flows developing behind bluff-bodies have been widely studied in the past because of their revelance in various industrial fields.
Obviously many bluff-body geometries are encountered in industrial applications. The drilling risers used in the offshore petroleum industry are, for instance, a realistic example of one of the most frequently studied bluff body flow configuration, the infinitely long circular cylinder. In sedimentology, spheres are quite often used to model particles sedimenting in a river bed. The well known Ahmed body, a three-dimensional object characterized by a rectangular-shaped cross section, is generally used as a car model in the automotive industry. These three examples are representative of two-dimensional wake flows and three-dimensional wake flows with reflectional symmetries.\\
For the previously mentionned industrial applications, the wake flows are in a turbulent state. Therefore, lots of studies 
have been dedicated to characterize the large- and small-scale fluctuations of the turbulent wakes. However recent experimental studies (\cite{Lawson2007},\cite{Herry2011},\cite{Grandemange2013}) have shown that three-dimensional turbulent wake flows additionaly exhibit bistability properties associated to the breaking of reflectional symmetries. Such bistability property is more commonly observed
in laminar flows developing in closed geometry (\cite{Golubitsky1988},\cite{Barkley2005}). Better understanding the laminar/turbulent transition of three-dimensional wake flows is thus still a major challenge today, which we address in this paper with linear stability theory.\\
The early stages of the laminar/turbulent transition scenario are now well known for wake flows behind two-dimensional geometries (\cite{Williamson1996a},\cite{Zdrakovitch1997}) and three-dimensional axisymmetric bodies (\cite{Thompson2001},\cite{Fabre2008}). 
However fewer studies have been dedicated to three-dimensional geometries exhibiting planar reflectional symmetries, such as the squared-back Ahmed body. Behind a two-dimensional circular cylinder, the two-dimensional steady wake flow is known to first 
bifurcate towards a two-dimensional time-periodic state when increasing the Reynolds number. This two-dimensional time-periodic flow subsequently bifurcates towards a three-dimensional time-periodic state when further increasing the Reynolds number \cite{Williamson1996b},\cite{Barkley1996},\cite{Thompson1996}. In this scenario the onset of unsteadinesses in the flow occurs prior to the appearance of three-dimensionality.\\
For three-dimensional axisymmetric geometries such as sphere and disks, the scenario is obviously different since the steady flow is necessarily three-dimensional. The axisymmetric steady wake flow first bifurcates towards a non-axisymmetric steady state preserving a symmetry plane. 
Subsequently a time-periodic wake flow emerges for which the planar symmetry is lost. Finally a planar-symmetry time-periodic flows is observed, the symmetry plane being here normal to the one  previously preserved \cite{Fabre2008}. In this scenario the breaking of the axisymmetry occurs prior to the onset of unsteadiness in the flow. \\
The fundamental difference between these two scenarios clearly lies in the nature of the first flow bifurcation. 
For two-dimensional geometries, the first bifurcation is a Hopf bifurcation, while for three-dimensional axisymmetric geometry the first bifurcation is a pitchfork bifurcation associated to the breaking of the rotational reflectional symmetry. 
Recent experimental investigations \cite{Grandemange2012} have shown that the transition scenario of wakes behind a three-dimensional bluff body of short aspect-ratio $L=1.35$  is close to the transition scenario behind axisymmetric bodies. More specifically, the steady flow preserving the planar symmetry of the bluff body bifurcates towards an asymmetric steady state breaking the planar symmetry. The first bifurcation is therefore a pitchfork bifurcation as in the case of axisymmetric body wakes. Nevertheless, observing such a permanent state requires to wait for a sufficiently long time since a planar-symmetric time-periodic flow is first observed in the experiments. The observation of such time-periodic transient state clearly suggests the existence of nearly unstable unsteady perturbations. Moreover, it is legitimate to expect that, for sufficiently large values of the aspect ratio, the destabilization scenario of a two-dimensional body is recovered, i.e. the first flow bifurcation is a Hopf bifurcation.\\
The effect of the spanwise extent on the wake flow transition behind various three-dimensional bodies has already been addressed in the past.
For instance, wakes transition behind finite-length cylinders with hemispherical ends or free ends 
has been investigated experimentaly in (\cite{Schouveiler2001},\cite{Provansal2004}) and numerically in (\cite{Sheard2008},\cite{Inoue2008})
with three-dimensional unsteady simulations of the Navier-Stokes equations.
For rectangular or elliptical cross-section plates, experimental studies (\cite {Kuo1967}, \cite{Kiya1999}, \cite{Kiya2001}) have been performed.
The present paper aims at investigating the wake flow transition behind rectangular flate-plates by means of linear stability analysis. 
Note that such geometry has been chosen because of its relevance with respect to path instabilities of 
three-dimensional object falling in a viscous fluid under the action of gravity \cite{Ern2012}. 
The first objective is to show that, for short aspect ratios, the flow first exhibits a pitchfork bifurcation, 
as for axisymmetric bodies. The second objective is to investigate how the transition scenario is modified when increasing the value of the aspect ratio.\\ 
The original aspect of the present paper, compared to previously mentionned studies, is the systematic use of global stability analysis of 
fully three-dimensional flows. The first global stability analysis has been performed on two-dimensional circular cylinder wakes \cite{Jackson1987}. 
An anti-symmetric unsteady global mode was found to be unstable, confirming that the first bifurcation is a Hopf bifurcation. 
Later on, global stability analysis of axisymmetric wake flows behind sphere and disks has been performed \cite{Natarajan1993},.
The steady axisymmetric flow gets first unstable to a steady mode of azimuthal wave number $m=1$ and then to an unsteady mode of same azimuthal wave number \cite{Meliga2009a}. 
More recently, various analyses have been proposed, based on the knowledge of unstable global modes and their adjoint global modes,
in order to determine the \textit{wavemaker} of the instability \cite{Giannetti2007}, to find the coefficients of the 
amplitude equation governing the non-linear evolution of the perturbation in the slow manifold \cite{Sipp2007}
or to design passive control strategy of unstable global modes based on sensitivity analysis \cite{Marquet2008a},\cite{Marquet2008b}.
In the case of a disk flows, the determination of two unstable global modes has been a pre-requisite to an elaborated weakly non-linear analysis
\cite{Meliga2009b} that gave the complete bifurcation diagramm of the wake. 
Up to now, only few global stability analyses have been performed on fully three-dimensional flow configuration, as for instance a jet in cross-flow (\cite{Bagheri2009}, \cite{Ilak2012}), probably because of the complexity of the underlying numerical problem. To overcome (disregard) this numerical complexity a strategy based on local absolute/convective stability analysis of steady flows in cross-stream planes were recently proposed 
and applied on the sphere wake \cite{Pier2008}. The stability of the fully three-dimensional wake flow developing 
behind the plates is here investigated by exploiting the spatial symmetries of the problem. This enables to partially reduce 
the computational cost of the problem.

The paper is organized as follows. The flow configuration and the global stability analysis are detailed in section \ref{sec:config}.
The computational methods, including the spatial discretization and the parallelization strategy, are presented in section \ref{sec:numerics}.
The results on the largest aspect ratio plate $L=6$ are detailed in section \ref{sec:largeaspectratio}.Finally section \ref{sec:aspectratio} is dedicated to the influence of the aspect ratio on the first flow bifurcation.

\section{Flow configuration and methodology}\label{sec:config}
\begin{figure}
\centering
\includegraphics[scale=.5]{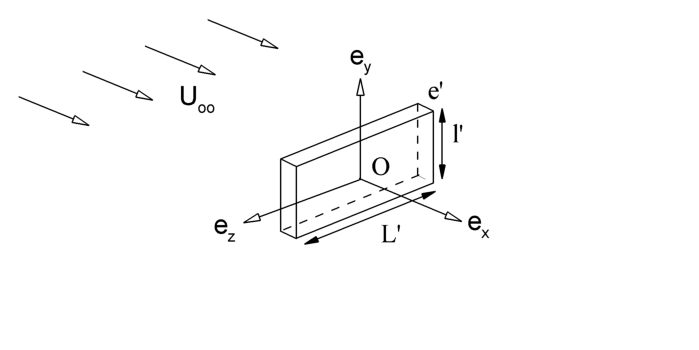} 
\vspace*{-1.0cm} \caption{Flow configuration}   
\label{fig:flowconf}       
\end{figure}

We investigate the incompressible flow of a Newtonian fluid of kinematic viscosity $\nu$ around three-dimensional plates defined by their length $L'$, width $l'$ and thickness $e'$. As seen in Figure \ref{fig:flowconf} a cartesian coordinate system $(O,\mathbf{e_{x}},\mathbf{e_{y}},\mathbf{e_{z}})$ is placed at the geometric center $O$ of the plate 
and the flow upstream to the plates is uniform of velocity $\mathbf{U}_{\infty}=U_{\infty} \mathbf{e}_{x}$. The plate width $l'$ and the upstream velocity $U_{\infty}$ 
are chosen as the reference length and velocity respectively and, from now, all quantities are normalized using these reference length and velocity. 
Three non-dimensional control parameters entirely define this flow configuration: two geometrical parameters, the thickness-to-width ratio denoted $e=e'/l'$ 
and the length-to-width ratio denoted $L = L'/l'$, and one flow parameter, the Reynolds number $\textit{Re}=U_{\infty} l' / \nu$.  In the present study 
the thickness-to-width ratio is fixed $e=1/6$ and the influence of the length-to-width ratio and Reynolds number on the wake transition
is studied in the parameter ranges $1 \le L \le 6$ and $40 \le \textit{Re} \le 150$. Since only the length-to-width ratio is varied here,
it will be referred in the following as the aspect ratio. Note that the aspect ratio $L=1$ corresponds to a squared plate while $L\rightarrow \infty$ 
would correspond to an infinitely long plate.\\ 

The incompressible three-dimensional flow is described by the velocity vector $\mathbf{u}(\mathbf{x})=(u,v,w)^{T}(\mathbf{x})$ and the scalar pressure field $p(\mathbf{x})$ which satisfy the non-dimensional unsteady Navier-Stokes equations    
\begin{eqnarray}\label{eqn:NS}
\partial_{t} \mathbf{u} + (\mathbf{u} \cdot \mathbf{\nabla}) \mathbf{u} + \nabla p - \frac{1}{\textit{Re}} \mathbf{\Delta} \mathbf{u} = 0 \;\;,\;\; \mathbf{\nabla} \cdot \mathbf{u} = 0
\end{eqnarray}
The linear stability analysis starts by decomposing the flow variables as the sum of a steady base flow $(\mathbf{U},P)(\mathbf{x})$ and an unsteady perturbation $(\mathbf{u}',p')(\mathbf{x},t)$, which is assumed to be infinitely small in magnitude compared to the base flow. To investigate the long-term stability, this perturbation is further decomposed into the normal mode form  $(\mathbf{u}',p')(\mathbf{x},t) = \sum_{k} (\mathbf{\hat{u}}_{k},\hat{p}_{k})(\mathbf{x}) \; e^{\lambda_{k} t} e^{\rm{i} \omega_{k} t} + \mbox{c.c.}$, i.e. as the sum of (complex) spatial structures $(\mathbf{\hat{u}}_{k},\hat{p}_{k})(\mathbf{x})$ whose individual temporal evolution is governed by the corresponding growth rate $\lambda_{k}$ and circular frequency $\omega_{k}$. By introducing such decomposition into the Navier Stokes equations (\ref{eqn:NS}) we obtain that the base flow satisfy the three-dimensional steady Navier-Stokes equations
\begin{eqnarray}\label{eqn:NSS}
(\mathbf{U} \cdot \mathbf{\nabla}) \mathbf{U} + \nabla P - \frac{1}{\textit{Re}} \mathbf{\Delta} \mathbf{U} = 0 \;\;,\;\; \mathbf{\nabla} \cdot \mathbf{U} = 0
\end{eqnarray}
while the global modes and their corresponding growth rate and frequency are the eigenvectors and eigenvalues of the following equations
\begin{eqnarray}\label{eqn:GEV}
(\lambda_{k}+\rm{i} \omega_{k}) \mathbf{\hat{u}}_{k} + (\mathbf{U} \cdot \mathbf{\nabla}) \mathbf{\hat{u}}_{k} + (\mathbf{\hat{u}}_{k} \cdot \mathbf{\nabla}) \mathbf{U} + \nabla \hat{p}_{k} - \frac{1}{\textit{Re}} \mathbf{\Delta} \mathbf{\hat{u}}_{k} = 0 \;\;,\;\; \mathbf{\nabla} \cdot \mathbf{\hat{u}}_{k} = 0
\end{eqnarray}
Assuming the eigenvalues are ordered by decreasing value of their growth rate, i.e. $\lambda_{0} \ge \lambda_1 \ge \lambda_2 \ge \cdots$, 
the long-term stability of the base flow is determined by the growth rate's sign of the leading eigenvalue $(\lambda_0,\omega_0)$. 
When $\lambda_0 < 0 $ the base flow is said to be globally stable, marginally stable when $\lambda_0 = 0$ and globally unstable when $\lambda_0 \ge 0$. 
If the angular frequency $\omega_0 = 0$, the flow is unstable to steady perturbations, while the flow is unstable to unsteady perturbations when $\omega_0 \neq 0$.

\begin{figure}
\begin{tabular}{lcrc}
(a) & & (b)&  \\
    & \includegraphics[scale=.4]{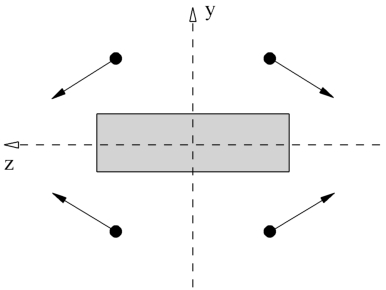} &  & \includegraphics[scale=.4]{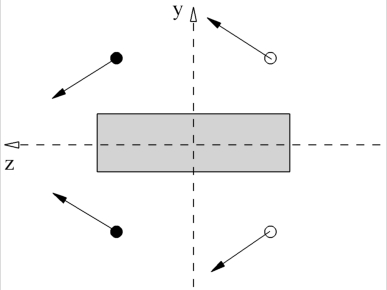} \\
(c) & & (d)&  \\
    & \includegraphics[scale=.4]{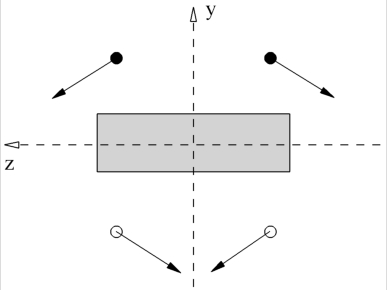} &  & \includegraphics[scale=.4]{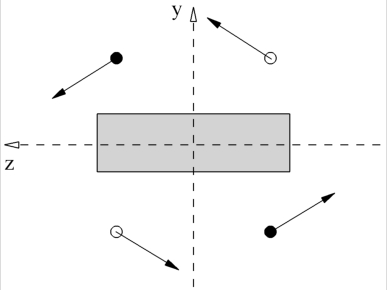}
\end{tabular}
\vspace*{-0.0cm} \caption{Illustration of the flow symmetry in a cross-stream plane. The flat plate is indicated in grey. The four flow symmetries  are (a) $(S_y,S_z)$ , (b) $(S_y,AS_z)$ , (c) $(AS_y,S_z)$ and (d) $(AS_y,AS_z)$. The arrows represent cross-stream velocity vectors $(v,w)$ while the black and white circles stand for streamwise velocities of opposite sign.}   
\label{fig:Symmetry}       
\end{figure}

The flat-plates exhibit two planar reflectional symmetries: one with respect to the plane $y=0$
and the other one with respect to the plane $z=0$. For sufficiently low values of the Reynolds number to be determined later, 
the flow is expected not only to be time-invariant but also to preserve the spatial symmetries of the body. 
Therefore it is legitimate to assume that the base flow satisfies the reflectional symmetries with respect to the 
planes $y=0$ and $z=0$, denoted in the following $S_{y}-$ and $S_{z}$-symmetry respectively. Due to the arbitrary orientation of the cartesian coordinate system shown in Figure \ref{fig:flowconf},
they will also be called top/bottom symmetry ($S_y$-symmetry) and left/right symmetry ($S_z$-symmetry).  
Figure \ref{fig:Symmetry}(a) provides a graphical illustration of these flow symmetries viewed in an arbitrary cross-stream plane $x$. The velocity vector $\mathbf{U}$ and pressure field $P$ of the base flow satisfy, for any point $\mathbf{x}=(x,y,z)$, 
the following symmetry relations
\begin{eqnarray}\label{eqn:symmetry}
(S_y) \;:\; (U,V,W,P)(x,-y,z) = (U,-V,W,P)(x,y,z)  \nonumber \\
(S_z) \;:\; (U,V,W,P)(x,y,-z) = (U,V,-W,P)(x,y,z) 
\end{eqnarray}
In other words, the $S_y$-symmetry states that the streamwise $U$, cross-stream $W$ velocities and pressure $P$ are even functions with respect to the variable $y$
while the cross-stream velocity $V$ is an odd function. For the $S_z$-symmetry, $U$,$V$ and $P$ are even functions with respect to $z$ and $W$ is an odd function.   
When increasing the Reynolds number, the flow is expected to break the time-invariance and/or the spatial symmetries (\ref{eqn:symmetry}) of the base flow. 
The time-invariance and spatial-symmetry breakings are determined by the frequency and spatial symmetries 
of the first unstable global mode. Four types of global modes exhibiting different combination of spatial symmetries may be expected. 
Hereinafter they will be referred as $(S_y,S_z)$, $(S_y,AS_z)$, $(AS_y,S_z)$ and $(AS_y,AS_z)$ global modes, 
where $AS$ stands for Anti-Symmetric. As an example, the velocity vector $\mathbf{\hat{u}}$ and pressure field $\hat{p}$ of an $(AS_y,AS_z)$ global mode satisfy the relations 
\begin{eqnarray}\label{eqn:antisymmetry}
(AS_y) \;:\; (\hat{u},\hat{v},\hat{w},\hat{p})(x,-y,z) = (-\hat{u},\hat{v},-\hat{w},-\hat{p})(x,y,z)  \nonumber \\
(AS_z) \;:\; (\hat{u},\hat{v},\hat{w},\hat{p})(x,y,-z) = (-\hat{u},-\hat{v},\hat{w},-\hat{p})(x,y,z)
\end{eqnarray}
which are graphically illustrated in Figure \ref{fig:Symmetry}(d). Apart from the physical interest to analyse the flow destabilization in terms of reflectional symmetry breaking, it is also useful to exploit those symmetries for saving computational ressources. The computations may be performed not on the whole physical space but on a restricted space, for instance $(y \ge 0,z \ge 0)$ which corresponds to the top-left quarter area in Figures \ref{fig:Symmetry}. Knowing the symmetries of the computed velocity and pressure fields, it is straightforward to reconstruct them on the whole physical space by using the definitions (\ref{eqn:symmetry}) and (\ref{eqn:antisymmetry}). 
The symmetries of the various flow fields are imposed in the computations via appropriate boundary conditions applied on the symmetry planes $y=0$ and $z=0$. 
For the $(S_y,S_z)$ base flow, the symmetric boundary conditions imposed on the symmetry planes are
\begin{eqnarray}\label{eqn:SySz}
(S_y):\; (\partial_y U,V,\partial_y W, \partial_y P)(x,0,z)=0 \nonumber \\ 
(S_z):\; (\partial_z U,\partial_z V,W, \partial_z P)(x,y,0)=0 
\end{eqnarray}
These symmetric boundary conditions state that the velocity component normal to the plane as well as the normal derivatives of the velocity vanishes.
For $(AS_y,AS_z)$ global modes , the anti-symmetric boundary conditions imposed on the symmetry planes are
\begin{eqnarray}\label{eqn:ASyAsz}
(AS_y):\; (\hat{u},\partial_y \hat{v},\hat{w},\hat{p} ) (x,0,z)=0 \nonumber \\ 
(AS_z):\; \hat{u},\hat{v},\partial_z \hat{w},\hat{p} ) (x,y,0)=0  
\end{eqnarray}
Obviously any type of global modes can be computed by using the appropriate combination of boundary conditions.

\section{Computational methods and convergence tests} \label{sec:numerics}
Global stability analysis is performed in two steps:
first finding a base flow which is a steady solution of the Navier-Stokes equations, 
then determining the stability of this base flow by looking 
for the leading global modes, i.e. modes of largest real part eigenvalues,
which are solutions of the eigenvalue problem (\ref{eqn:GEV}).
Generally speaking, the methods used to accomplish these two steps are 
classified as matrix-free methods or direct methods.
The former class of methods is based on the use of existing and efficient solvers 
which implement time-stepping techniques of the Navier-Stokes equations.
The determination of steady base flows and of their stability with matrix-free methods is explained in \cite{tuckerman2000}, \cite{bagheri2009}. Direct methods are used in the present study, as in \cite{sipp2010}, where the stability of two-dimensional base flows have been performed. To efficiently compute the stability of three-dimensional base flows, a fully parallel strategy is required and detailed in this section. The spatial discretization and the iterative methods used to solve the two problems are first recalled. Finally convergence tests are carried out to assess the accurary of base flow and eigenvalues with respect to the computational size box and the mesh refinement.  

\subsection{Spatial discretization and numerical methods}
All of the partial differential equations involved in the present study are discretized in space using a continuous Galerkin finite element discretization.
Weak formulations of the equations (\ref{eqn:NSS}) and (\ref{eqn:GEV}) are first determined and then spatially discretized on meshes composed of tetrahedra. 
The local polynomial bases chosen for the weight and test functions in the Galerkin projection are of second-order for the velocity components and of first order for the pressure. This choice ensures the numerical stability of the spatial discretization for incompressible flows since it satisfies the Ladyzhenskaya Babuska Brezzi (LBB) condition. 
\begin{figure}
\begin{tabular}{ll}
(a) & (b)  \\
 \includegraphics[scale=.3]{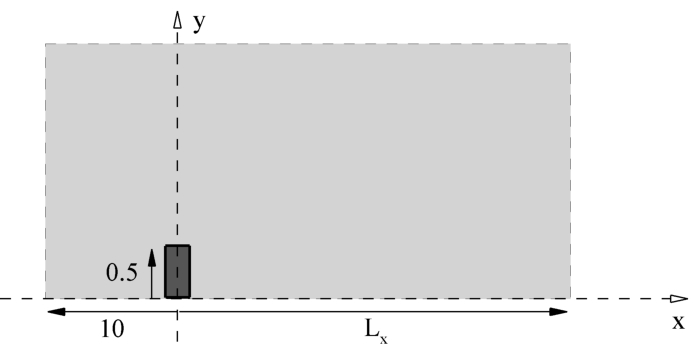} & \includegraphics[scale=.3]{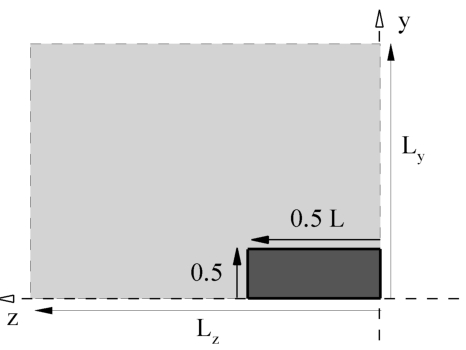} 
\end{tabular}
\vspace*{-0.0cm} \caption{The computational box is a quarter of the physical space. (a) Side view. (b) Rear view. The plate is in dark gray. The light gray
shows the fluid domain.}   
\label{fig:ComputationalBox}       
\end{figure}

The mesh generation and assembly of the matrices obtained by discretization of the weak formulations are done within the software FreeFem++ \cite{freefem} freely available on the website \textsl{www.freefem.org/ff++/}. 
Figure \ref{fig:ComputationalBox} shows the computational box used in the present study. As explained in the previous section the symmetries of the problem are used to reduce the computational box to the positive quadrant $(y\ge0,z\ge0)$, as seen in Figure \ref{fig:ComputationalBox}(b). The distance of the inlet 
to the center of the plate $O$ is fixed to $10$ in the following. It has been checked that this choice does not affect the results. 
The distance of the outlet, side and top boundaries to the center of the plate are respectively denoted $L_x$, $L_y$ and $L_z$. 
Tests on the convergence of the results with respect to the size of the computational box defined by this distances are reported below.
Once the size of the computational box is chosen an unstructured mesh is generated with the freely available library \textit{TetGen} which is interfaced with FreeFem++. It creates a tetrahedral mesh composed of $n_{e}$ tetrahedra and $n_{v}$ vertices. Table \ref{tab:mesh} displays the characteristics of four meshes used in the following. The number of degree of freedom $n_{dof}$ corresponds to the size of a discretized solution, denoted now $\mathbf{Q}$ for the base flow,
and counts for the size of the three velocity components and of the pressure. \\

The first problem is a steady non-linear equation (\ref{eqn:NSS})  which is here solved using a Newton method. This iterative method is briefly explained. The discretized solution $\mathbf{Q}_{j+1}$ at the $j+1$ iteration of the process 
is searched in the form $\mathbf{Q}_{j+1}=\mathbf{Q}_{j}+\mathbf{\delta Q}_{j}$ where $\mathbf{Q}_{j}$ is the known solution at the previous iteration $j$
while $\mathbf{\delta Q}_{j}$ is the unknown correction. Introducing this decomposition into the nonlinear equation (\ref{eqn:NSS}), one obtains the correction equation which writes in a discrete setting 
\begin{eqnarray}\label{eqn:Newton}
\mathbf{A}_{j} \cdot \mathbf{\delta Q}_{j} = \mathbf{R}(\mathbf{Q}_{j})
\end{eqnarray}
where $\mathbf{A}_{j}$ is called the Jacobian matrix and  results from the spatial discretization of the linearized Navier-Stokes equations around the solution at the previous step $\mathbf{Q}_{j}$. $\mathbf{R}(\mathbf{Q}_{j})$ is the residual vector evaluated at the iteration $j$.
The matrix $\mathbf{A}_{j}$ is a large matrix of size $n_{dof} \times n_{dof}$ but in a finite element setting this matrix is sparse. 
The sparsity of the matrix can be assessed by the ratio $s$ between the number of non-zero elements of the matrix $N_z$ over the number of degrees of freedom $n_{dof}$. Typical values of $N_z$ and $s$ are given in Table (\ref{tab:mesh}). Compared to similar computations but in a two-dimensional setting, the sparsity of the matrix 
is large. Therefore the memory needed to store such matrices is non negligible, as it can be also seen in Iable \ref{tab:mesh}. 
The iterative process is stopped once the $L_2$-norm of the residual is less than a tolerance fixed to $10^{-12}$ in the following. Typically, $5$ to $6$ iterations are needed to converge the Newton algorithm. \\

\begin{table}
\centering
\begin{tabular}{rcrcrcc}
Mesh \vline & $n_{e}(\times 10^3)$ & $n_{v}(\times 10^3)$ \vline & $n_{dof}(\times 10^6)$ & $N_{z}(\times 10^6)$ \vline & $s$ & Memory (Mb) \\
\hline
$M_1$ \vline & $291$ & $58$ \vline & $1.36$ & $111$ \vline & $70$ & $1776$ \\ 
$M_2$ \vline & $408$ & $81$ \vline & $1.90$ & $166$ \vline & $87$ & $2656$ \\ 
$M_3$ \vline & $588$ & $113$ \vline & $2.70$ & $238$ \vline & $88$ & $3808$ \\
$M_4$ \vline & $805$ & $150$ \vline & $3.61$ & $328$ \vline & $90$ & $5248$  \\
\end{tabular}
\caption{Characteristics of four meshes used for testing the refinement convergence. $n_e$: number of elements (tetrahedra) in the mesh; $n_v$: number of vertices; $n_{dof}$: number of degree of freedom of a discretized solution. $N_z$: number of non-zero entries in the sparse matrix representing the linearized Navier-Stokes equations at one step of the Newton method. Sparsity $s$ and memory needed to store such matrix. 
Memory is expressed in Megabytes.}
\label{tab:mesh}
\end{table}

The stability of the steady solution is then investigated by solving the generalized eigenvalue problem (\ref{eqn:GEV}). 
The spatial discretization of this generalized eigenvalue problem leads to the matrix equation
\begin{eqnarray}\label{eqn:GEVD}
\mathbf{A} \cdot \mathbf{\hat{q}}_{k} = \sigma_{k} \mathbf{B} \cdot \mathbf{\hat{q}}_{k}
\end{eqnarray}
where $\sigma_{k}= \lambda_{k} + \rm{i} \omega_{k}$ is a complex eigenvalue and $\mathbf{\hat{q}}_{k}$ is a complex eigenvector of size $n_{dof}$.
$\mathbf{A}$ and $\mathbf{B}$ are usually called the jacobian and mass matrix respectively. Only the largest real part    
eigenvalues are of interest to determine the flow stability, as explained in the previous section. To obtain those particular eigenvalues a shift-and-invert strategy is used. It consists in solving, instead of (\ref{eqn:GEVD}), the following problem  
\begin{eqnarray}
\left( \mathbf{A} - \sigma_{s} \mathbf{B} \right)^{-1} \mathbf{B} \cdot \mathbf{\hat{q}}_{k} = (\sigma_{k}-\sigma_{s})^{-1} \mathbf{\hat{q}}_{k}
\end{eqnarray}
where $\sigma_{s}$ is a complex number called a shift. The eigenvalues are computed using a variant of the Arnoldi method, called the Implicitly Restarted Arnoldi Method, and implemented in the library ARPACK \cite{arpack} and PARPACK, which is 
the implementation for distributed memory parallel architecture. The later has been used with the shift-and-invert mode. This mode requires to implement two functions, one performing the matrix-vector product 
\begin{eqnarray}
\mathbf{z}_{out} \leftarrow \mathbf{B} \cdot \mathbf{z}_{in} 
\end{eqnarray}
and one giving the solution of the linear system 
\begin{eqnarray}\label{eqn:LSC}
\left( \mathbf{A} - \sigma_{s} \mathbf{B} \right) \mathbf{z}_{out} = \mathbf{z}_{in}
\end{eqnarray}
Note that these operations are repeated as many times as the number of iterations needed by the Arnoldi algorithm 
to converge towards the number of eigenvalues requested by the user. The matrix-vector product is the cheapest operation 
and the bottleneck of the method clearly lies in the resolution of the linear system.
The strategy for the parallelization of this linear system is detailed in the next paragraph.

\subsection{Parallelisation strategy and performance}
A parallelization strategy is needed to efficiently solve the two linear systems (\ref{eqn:Newton}) and (\ref{eqn:LSC}). A direct method is used to invert these large scale linear problems. It is based
on an explicit construction/assembly of the matrices $\mathbf{A}_j$, $\mathbf{B}$ and $\mathbf{A}-\sigma_s \mathbf{B}$ and an inverting phase, composed of an advanced lower-upper factorization method of the matrices $\mathbf{A}_j$ and $\mathbf{A} - \sigma_{s} \mathbf{B}$ and followed by two fast triangular system resolution. During the base flow computation, the assembly and factorization phases are repeated at each iteration of the Newton method. For an eigenvalue computation, these phases are done once for all, while the solving phase is repeated as many times as the number of iterations needed for the Arnoldi method to converge.

The parallelization of the assembly phase is based on a non-overlapping partitioning of the mesh. 
The mesh is first built on one processor with the library \textit{TetGen} and then splitted into $N_{MPI}$ non-overlapping subdomains $\Omega_{i}$. To that aim the library \textit{METIS} \cite{metis} 
is used as a graph partitionner to properly balance the number of elements among each processor. Note that this operation is serial and its parallelization is not really needed since the computational time associated to this operation is negligible. Then each $MPI$ process $i$ assembles the sub-matrix $A_{i}$ which corresponds to the discretization of the problem of interest on 
the subdomains $\Omega_{i}$. The matrix is said to be in a \textit{distributed} assembled format. 
Note that the discretized matrix on the full domain $\Omega$ is never assembled on one processor. The parallelization of the assembly phase has two advantages. Firstly, it distributes the memory cost of the matrix assembly on many cores. Secondly, it speeds up this phase by the number of $MPI$ processes. Indeed the construction of the matrices $A_{i}$ is entirely parallel and no information needs to be exchanged between the $MPI$ processes. Once the assembly is finished, each process writes its sub-matrix on the disk.
 
The inverting phase is performed with the MUltifrontal Massively Parallel sparse direct Solver (\textit{MUMPS} \cite{mumps}). The distributed assembled format of this library is used, thus each $MPI$ process reads the matrix $A_i$ stored on the disk. Inversion with most direct solvers is in fact made of three steps: analysis phase, factorization phase and forward/backward triangular solving phase. The well-know advantage of direct methods over iterative methods is their computing-time efficiency and robustness.
On the other hand they are also well known to consume a lot amount of memory. Briefly, the memory needed to store the lower $L$ and upper $U$ matrices 
obtained during the factorization phase can be much larger than twice the memory needed for $A$. This is due to the sparsity of $L$ and $U$ that can be much larger than the sparsity of $A$. 
\begin{table}
\centering
\begin{tabular}{rccccrr}
\\
\vline & & Memory & (Mb)&  & \vline & Time (s) \\
\hline
Mesh     \vline &  Maximum & Average & $8$ cores & Total & Occupancy (\%) \vline &  \\  
\hline
$M_1$ \vline & $1629$ & $-$ & $-$ & $-$ & $-$ \vline & $126$ \\ 
$M_2$ \vline & $2468$ & $1827$ & $7308$ & $58 368$ & $20$ \vline & $180$ \\ 
$M_3$ \vline & $4032$ & $2902$ & $11608$ & $92 864$ & $32$ \vline & $410$ \\ 
$M_4$ \vline & $5777$ & $4478$ & $17912$ & $143 296 $ & $49$ \vline & $742$ \\ 
\end{tabular}
\caption{Memory consumption in Megabytes (Mb) with respect to the mesh refinement during base flow computations. 
Memory informations as given by the library MUMPS during the factorization phase. Maximum stands for the maximum memory used by one of the MPI processes. 
Average indicates is the total memory, given by the column Total, divided by the number of MPI processes.
Computations have been performed here using $64$ cores with $32$ MPI processes and $2$ OpenMP threads. The column $8$ cores corresponds to the memory used by $1$ computational node of $8$ cores. It is computed as the average memory multiply by the number of $MPI$ process on $1$ node, $4$ in the present case.}
\label{tab:perfo1}
\end{table}

Table \ref{tab:perfo1} displays the memory and computational time needed during the factorization phase of one iteration of the Newton method. The four cases presented in Table \ref{tab:mesh} are analyzed. Computations have been performed on a SGI cluster 
(AltiX ICE 8200 EX Nehalhem) using $64$ cores with $32$ MPI processes and $2$ OpenMP threads. The mesh is partitioned into a number of 
sub-meshes equal to the number of MPI processes, i.e. $32$ for the described computations. The columns entitled Maximum, Total and Average  
corresponds respectively to the maximum memory used by one of the $MPI$ process, the total memory, and the total memory divided by the number of MPI processes. 
These informations are given by MUMPS during the factorization phase. Because the cluster is made of computational distributed-memory nodes composed of $8$ sharing-memory cores, the column $8$ cores has been added to count the average memory used per node. It is computed as the average memory multiply by the number of $MPI$ process on one node. For the described computation, $4$ MPI processes are used on each of the $8$ used nodes. It should be compared to the maximum memory available per node, equal to $36000$ Mb one the present machine. This comparison is made on the next column entitled Occupancy which gives the average percentage of occupancy of nodes during this factorization phase. For the finest mesh $M_4$, the size of the linear system is $3.61$ millions of degrees of freedom and requires a total memory of $143$ Gb, 
which is $27$ times the memory needed for the matrix storage. This is a huge amount of memory but it still only corresponds to $49\%$ of the total memory that might be used. 
The last column gives the elapsed time during the factorization phase, i.e. the time as experienced by the user. Obviously it depends on the number of cores 
used for the computation. For the largest memory-consuming case ($M_4$), the factorizatipon phase last about $12$ minutes when using $64$ cores. 
This corresponds to about $60 \%$ of the total time, $20 \%$ being consumed in each of the assembly and solve phases. 
Solving the linear sytem (\ref{eqn:LSC}) for the eigenvalue computation requires twice as much memory as when solving the linear system (\ref{eqn:Newton})
for computing the base flow, because of the complex arithmetics needed for the matrix $\mathbf{A} - \sigma_{s} \mathbf{B}$. 

\subsection{Convergence tests}
The influence of the mesh refinement on the numerical results is here investigated. 
The results of these tests are now described, for the base flow and stability computations. 
\begin{table}
\centering
\begin{tabular}{rlccr}
\\
Mesh \vline & $L_{b}$ & $H_{y}$ & $C_{D}$ & $U_{min}$ \\ 
\hline
$M_{1}$ \vline & $4.468$ & $1.932$ & $0.592$ & $-0.45138$ \\
$M_{2}$ \vline & $4.470$ & $1.932$ & $0.592$ & $-0.45123$ \\
$M_{3}$ \vline & $4.472$ & $1.934$ & $0.592$ & $-0.45123$ \\
$M_{4}$ \vline & $4.474$ & $1.934$ & $0.592$ & $-0.45128$ \\
\end{tabular}
\caption{Convergence test of the base flow solution with respect to the mesh refinement. The length $L_b$ and height $H_y$ of the recirculation region, 
the drag coefficient $C_{D}$ and the maximal backflow velocity are displayed. Control parameters $\textit{Re}=100$, $L=2.5$.}
\label{tab:convbaseflow}
\end{table}

Table \ref{tab:convbaseflow} shows results for the mesh refinement tests on the base flow.
The base flow accuracy is aleviated by examining four quantities: the drag coefficient $C_{D}$, the length $L_{b}$ and height $H_y$ 
of the recirculation region and the maximal backflow velocity in this recirculaiton region $-U_{min}$. 
Note that all of these quantities will be better define latter. As seen in Table \ref{tab:convbaseflow}, 
all of these quantities are very weakly modified when the mesh is refined.

\begin{table}
\centering
\begin{tabular}{rlrcr}
\\
Symmetry \vline &           & $(AS_y,S_z)$ \vline           &   & $(S_y,AS_z)$ \\
\hline 
Mesh     \vline & $\lambda$ & $\omega$      \vline & $\lambda$ & $\omega$ \\ 
\hline
$M_{1}$ \vline & $-0.009112$ & $0.57944$ \vline & $-0.0015120$ & $0.30599$  \\
$M_{2}$ \vline & $-0.008423$ & $0.58011$ \vline & $-0.0013625$ & $0.30598$  \\
$M_{3}$ \vline & $-0.008039$ & $0.58030$ \vline & $-0.0012765$ & $0.30585$  \\
$M_{4}$ \vline & $-0.008084$ & $0.58028$ \vline & $-0.0012155$ & $0.30584$  \\
\end{tabular}
\caption{Convergence test for two leading eigenvalues with respect to the mesh refinement. Control parameters $\textit{Re}=100$, $L=2.5$.}
\label{tab:convstab}
\end{table}

The effect of the mesh refinement on two modes with different symmetries is displayed in Table \ref{tab:convstab}.
The eigenvalues associated to the leading $(AS_y,S_z)$ and $(S_y,AS_z)$ global modes are shown for the four meshes.
When comparing the results obtained for the meshes $M_3$ and $M_4$, the variation of the growth rate and circular frequency 
of the $(AS_y,S_z)$ mode is less than $1 \%$. For the $(S_y,AS_z)$ mode, the variation is slightly larger for the growth rate, around $4 \%$.
This indicates that results shown in the following can be considered as independent of the mesh refinement.

\section{The plate of largest aspect-ratio $L=6$}\label{sec:largeaspectratio}
The global stability of the flow around a plate of aspect ratio $L=6$ is investigated in this section.
This is the largest aspect-ratio considered in the present study and the cases of smaller aspect-ratios would be considered in the next section.
The base flow and its stability are first described at the specific Reynolds number $\textit{Re}=60$. 
Then, by varying the Reynolds number is then varied in the range $40 \le \textit{Re} \le 120$,  and successive destabilization of global modes are identified.
\begin{figure}
\begin{minipage}{0.4\textwidth}
\begin{tabular}{lc}
(a) & \\
& \includegraphics[scale=.5]{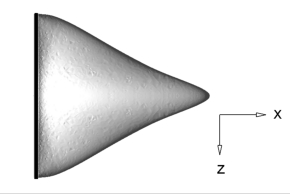} \\
(b) & \\
& \includegraphics[scale=.5]{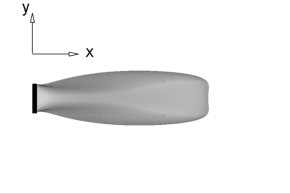} 
\end{tabular}
\end{minipage}
\begin{minipage}{0.6\textwidth}
\begin{tabular}{lc}
(c) & \\
& \includegraphics[scale=.5]{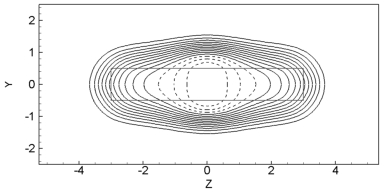} \\
(d) & \\
& \includegraphics[scale=.5]{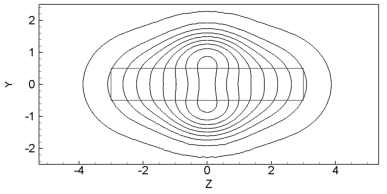} 
\end{tabular}
\end{minipage}
\vspace*{-0.0cm} \caption{Base flow around a flat-plate of length-to-width ratio $L=6$ at the Reynolds number $\textit{Re}=60$. 
Top (a) and side (b) views of the isosurface of zero streamwise velocity delimiting the recirculation region in the wake of the flat-plate.
Isocontours of streamwise velocity in cross-stream planes (c) $x=2.5$ and (d) $x=8$. The lines corresponds to decreasing values from $1$ to $-0.3$ 
by increments of $-0.1$. The dashed line stand for negative values. The rectangle in (c-d) is the flat-plate.}   
\label{fig:baseflow1}       
\end{figure}

\subsection{Global stability of base flow for $\textit{Re}=60$}
The base flow obtained for the Reynolds number $\textit{Re}=60$ is displayed in 
Figure \ref{fig:baseflow1}. The wake is characterized by a large three-dimensional recirculation region where the streamwise velocity 
is negative $U(\mathbf{x})<0$. This region is separated from the flow oriented downstream 
by the surface where the streamwise velocity is strictly equal to zero. This surface is depicted in Figures \ref{fig:baseflow1}
(a) and (b). The top view shown in Figure \ref{fig:baseflow1}(a) clearly indicates that the size of the recirculation region 
shrinks in the spanwise direction as compared to the length of the plate. Oppositely 
the size of the recirculation region extends in the vertical direction when compared to the width of the plate, 
as seen in the side view in Figure \ref{fig:baseflow1}(b). To further quantify the topology of the recirculation region, 
the length $L_{b}$ and height $H_{b}$ of this bubble are defined as follows. The length $L_{b}$ is the streamwise station 
where the streamwise velocity on the middle axis vanishes, i.e. $U(L_{b},0,0)=0$. In the present case it is equal to $L_{b}=6.33$.
The height $H_b$ is defined as twice the maximal vertical coordinate of points on the recirculation line in the plane $z=0$. 
It is here equal $H_{y}=2.01$, i.e. about twice the width of the plate. In addition to characterize the topology of the recirculation region, 
the maximal backflow velocity is determined to quantify the intensity of the flow recirculation.  This velocity, denoted 
$-U_{min}$, is defined as the opposite the minimal streamwise velocity on the central axis $U_{min}$. In the present case the maximal backflow velocity is equal to 
to $-U_{min}=0.44$. The streamwise velocity defect in the wake of the plate is also visible in Figures \ref{fig:baseflow1}(c-d) which show isocontours of the streamwise velocity in two cross-stream planes. The first one, depicted in Figure \ref{fig:baseflow1}(c), corresponds to the station $x=2.5$ and is a representative example of the near wake.
The second one, depicted in Figure \ref{fig:baseflow1}(d), corresponds to the station $x=8$ and is a representative example of the far wake.
\begin{figure}
\centering
\begin{tabular}{ll}
(a) & (b) \\
\includegraphics[scale=.25]{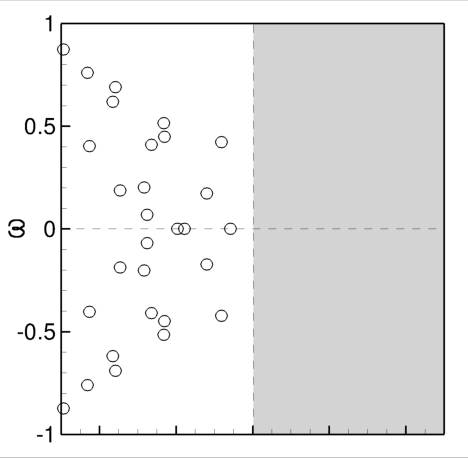} & \includegraphics[scale=.25]{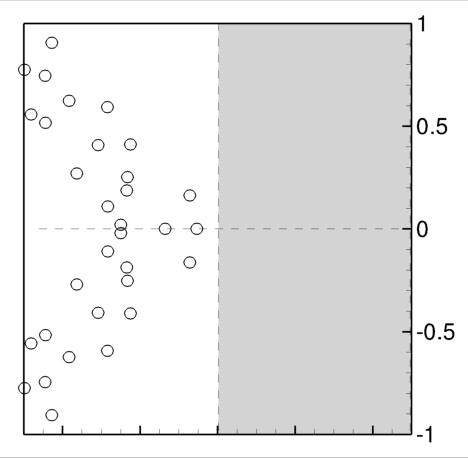} \\
(c) & (d)  \\
\includegraphics[scale=.25]{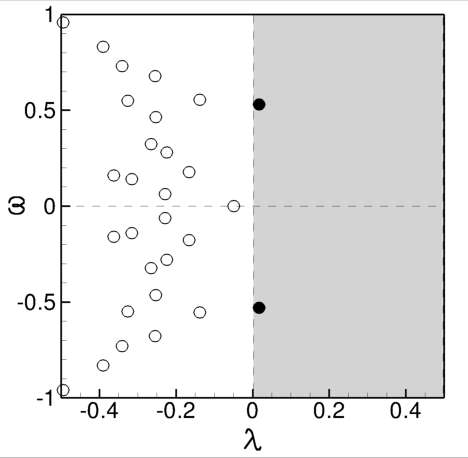} & \includegraphics[scale=.25]{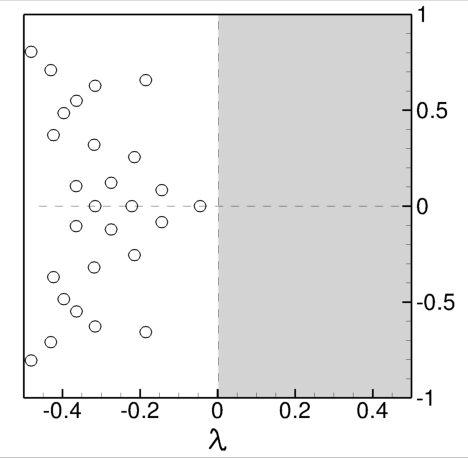}
\end{tabular}
\vspace*{-0.0cm} \caption{Eigenvalue spectrums of the base flow for $L=6$ and $\textit{Re}=60$ (see Figure \ref{fig:baseflow1}). The symmetry of the associated spatial structures is (a) $(S_y,S_z)$ , (b) $(S_y,AS_z)$ , (c) $(AS_y,S_z)$ and (d) $(AS_y,AS_z)$. The unstable half-plane is in grey.}   
\label{fig:Spectres}       
\end{figure}
Even though the plate is rectangular, the streamwise velocity isocontours quickly recover an elliptical-like shape as seen in Figure \ref{fig:baseflow1}(c). 
In the near wake the major axis of those elliptical-like contours is aligned with the length of the plate. In the far wake the shape of the isocontours 
still looks like an ellipse but, for small values of the velocity, the major axis is now aligned with the width of the plate. 
A similar effect has been observed in an experimental study about the wake of elliptical plates \cite{Kuo1967}.
The aspect ratios in their study are comparable to those investigated in the present study but their Reynolds numbers are much larger ($ 8 \, 10^3 \le \textit{Re} \le 7 \, 10^4 $) so that 
the wake is turbulent. In another experimental investigation \cite{Kiya1999}
for Reynolds number  $\textit{Re}=2.10^4$, a so-called axis switching 
on the mean and fluctuating streamwise velocity has been noticed,
occuring around $x=4$ ($x=4.5$) for elliptic plates of aspect ratio $L=2$ and $L=3$. 

They argued that this axis switching was provoked by the shape of the hairpin-like structures shed in the wake. 
The present results show that this phenomenom is also visible in the wake of the base flow for low Reynolds number and therefore independently of the existence of 
any fluctuations. This does not preclude that the shedding of specific vortical structures may reinforce this phenomenom as argued in \cite{Kiya1999}. \\ 
A further examination of Figures \ref{fig:baseflow1}(c) and (d) clarify the shape of the top/bottom and left/right shear layers. 
The top/bottom shear layers are characterized by large variations of the streamwise velocity in the vertical direction, i.e. strong velocity gradient $|\partial_{y} U|$.
On the other hand the left/right shear layers are characterized by large streamwise velocity gradient in the spanwise direction $|\partial_{z} U|$. 
For the plate of aspect ratio $L=6$, the left/right shear layers are much weaker than the top/bottom shear layers, but when decreasing 
the aspect ratio it is expected that the left/right shear layer get stronger. The coexistence of those two shear layers clearly suggests that 
a competition between two wake instabilites should occur in this flow: the instability induced by the interaction of the top and bottom shear layers 
and the instability induced by the interaction of the left and right shear layers. The frequencies associated to these wake instabilities are expected to scale 
on different lengths, which are approximatively the width and length of the plate.\\ 

The linear stability of this base flow is now addressed using the global stability analysis exposed in section 2.
The eigenvalue spectra corresponding to the four types of symmetry combination are depicted in Figure \ref{fig:Spectres}.
For instance, the spectrum of eigenvalues associated to $(S_y,S_z)$ global modes is depicted in Figure \ref{fig:Spectres}(a).
All of the eigenvalues lie in the stable left half-plane except for a pair of complex eigenvalues which lies in the unstable right half-plane in Figure \ref{fig:Spectres}(c). 
For the plate of aspect ratio $L=6$, the first flow bifurcation is thus a Hopf bifurcation which breaks the $S_{y}$-symmetry of the base flow 
since the corresponding unstable global mode satisfies the spatial symmetry $(AS_{y},S_{z})$. The angular frequency of this mode is $\omega=0.529$,
i.e. a non-dimensional frequency $f=\omega/2 \pi = 0.084$. The flow would bifurcate from the steady base flow towards a non-linear time-periodic flow.
The determination of this attracting state is out of the scope of the present study. But the frequency of the global mode is a linear approximation of the 
the fundamental frequency characterizing this attracting state, even if it is expected to be a poor approximation especially far from the critical Reynolds number (\cite{Sipp2007}).
\begin{figure}[htbp]
\begin{tabular}{rl}
(a) & \\
& \includegraphics[scale=.5]{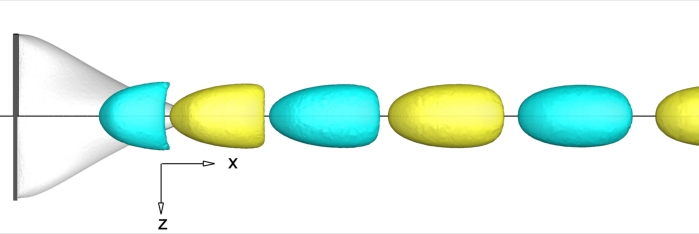} \\
(b) & \\
& \includegraphics[scale=.5]{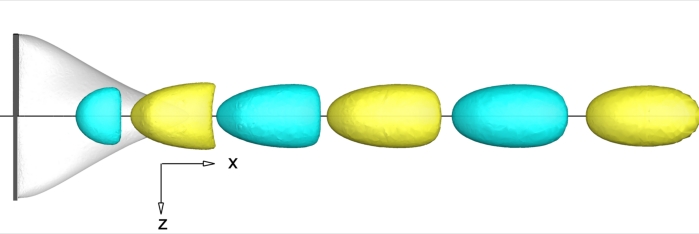} \\
(c) & \\
& \includegraphics[scale=.5]{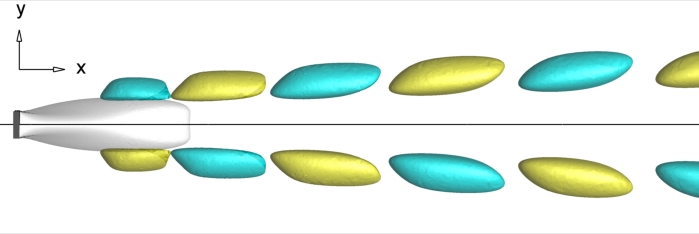} \\
(d) & \\
& \includegraphics[scale=.5]{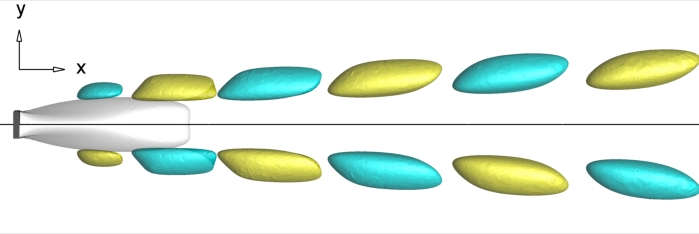} 
\end{tabular}
\vspace*{-0.0cm} \caption{The unstable $(AS_y,S_z)$ global mode for $\textit{Re}=60$ and $L=6$. The associated eigenvalue is shown by the black circle  
in Figure \ref{fig:Spectres}(c). Isosurfaces of positive (yellow) and negative (blue) streamwise velocity superimposed onto the base-flow recirculation surface
displayed in grey. Top (a-b) and side (c-d) views of the real and imaginary parts. Figures should be finished.}
\label{fig:Mode1}       
\end{figure}
The spatial distribution of the unstable global mode is depicted in Figure \ref{fig:Mode1} with isosurfaces of positive (yellow) and negative (blue) 
streamwise velocity. Spatial structures of alternating sign in the streamwise direction are clearly visible in the top views displayed in Figure \ref{fig:Mode1}
(a-b). The largest absolute values of streamwise velocity of this mode are reached in the minor plane $z=0$ and in the top and bottom shear layers of the base flow, 
as seen in Figure \ref{fig:Mode1} (c-d). The side views displayed in these figures also shows the alternate sign of the perturbation streamwise velocity between 
the top and bottom shear layers. Finally a comparison of the real and imaginary parts of the global mode, respectively shown 
in Figures \ref{fig:Mode1}(a-c) and (b-d), indicate that they are spatially out of phase. All of these features suggest that 
the wake pattern of the time-periodic flow is an alternate shedding of vortical structures between the top and bottom shear layers.
A more detailed investigation of this strucutre should reveal that it looks like a double sided hairpin-like vortices.

\subsection{Reynolds number effects}

\begin{figure}
\centering
\begin{tabular}{ll}
(a) & (b) \\
\includegraphics[scale=.4]{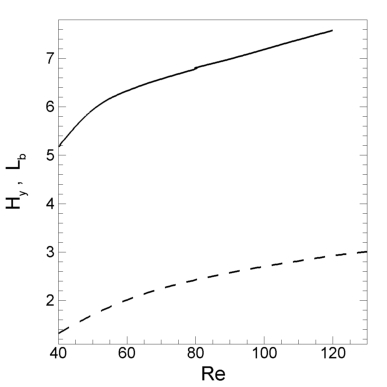} & \includegraphics[scale=.4]{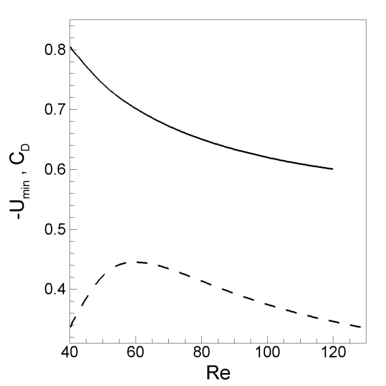} \\
\end{tabular}
\vspace*{-0.0cm} 
\caption{Effect of the Reynolds number on the base flow characteristics. (a) Length $L_{b}$ (solid line) and heigth $H_y$ (dashed line) of the recirculation region. (b) Drag coefficient $C_{D}$ (solid line) and maximal backflow velocity $-U_{min}$ (dashed line). Parameter: $L=6$.}   
\label{fig:BaseFlowReynolds}
\end{figure}

The Reynolds number is now varied to study its effect onto the base flow characteristics and the stability properties.
The variation of the base flow topology is first examined. The length and height of the recirculation regions is displayed in Figure \ref{fig:BaseFlowReynolds}(a) 
as a function of the Reynolds number. They both show a monotone increase when increasing the Reynolds number. For the smalest values of Reynolds number $\textit{Re}=40$, 
the height of the recirculation region $H_y$ is already greater than $1$ indicating that the extent of the recirculation region in that direction is larger than the width of the plate. 
The drag depicted by the solid line in Figure \ref{fig:BaseFlowReynolds}(b) shows a monotonic decrease when increasing the Reynolds number. Finally the evolution of the maximal backflow, displayed by the dashed line velocity in Figure \ref{fig:BaseFlowReynolds}(b),  is more complicated. The backflow velocity increases for low value of the Reynolds number, reaches a peak value at $\textit{Re} \le 60$, and finally decreases for larger values of the Reynolds number. \\
  
\begin{figure}
\centering
\begin{tabular}{ll}
(a) & (b) \\
\includegraphics[scale=.4]{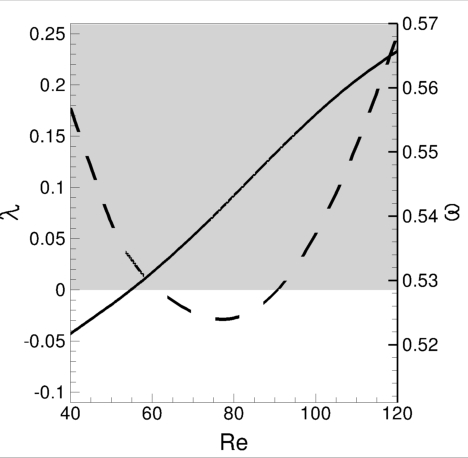} & \includegraphics[scale=.4]{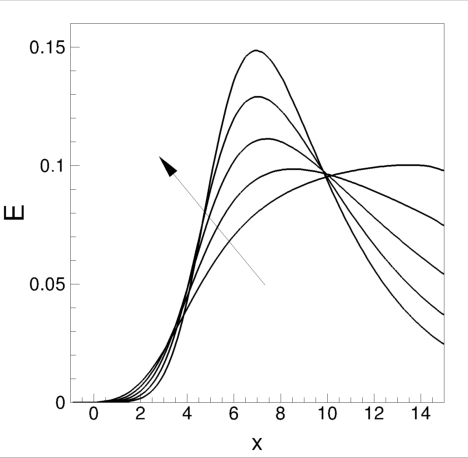}    
\end{tabular}
\vspace*{-0.0cm} 
\caption{Effect of the Reynolds number on the leading $(AS_y,S_z)$ global mode . (a) Growth rate and (solid line) and angular frequency (dashed line) 
of the unstable $(AS_y,S_z)$ mode as a function of the Reynolds number. (b) Kinetic energy of the global mode integrated in a cross-stream plane 
as a function of the streamwise station $x$ of this plane. The lines correspond to various values of the Reynolds numbers from $\textit{Re}=50$ to $\textit{Re}=90$ by step of $10$.}   
\label{fig:FirstModeRe}
\end{figure}

The growth rate and angular frequency of the $(AS_y,S_z)$-symmetric global mode described in the previous paragraph  
are shown in Figure \ref{fig:FirstModeRe}(a) as a function of the Reynolds number. The growth rate, depicted by the solid line in the 
Figure, changes sign for $\textit{Re}_{c1} \sim 55$, which determines the critical Reynolds number of the first flow bifurcation. This is a Hopf bifurcation 
and the angular frequency of the global mode is depicted in the same figure by the dashed line. The angular frequency is around $\omega_{c1} \sim 0.53$ at the critical Reynolds number. For larger Reynolds number it first decreases before increasing for $\textit{Re} \ge 80$. The evolution of the global mode structure with the Reynolds number is examined using the cross-stream kinetic energy of the mode.
It is defined as 
\begin{eqnarray*}
E(x) = \int_{\Gamma(x)} \left( \hat{u}^{*} \hat{u} + \hat{v}^{*} \hat{v} + \hat{w}^{*} \hat{w} \right) \; dy \, dz    
\end{eqnarray*}
where $\Gamma(x)$ is a cross-stream plane located at the station  $x$. 
This quantity is shown in Figure \ref{fig:FirstModeRe}(b) as a function of the cross-stream position $x$ of the plane 
and for various values of the Reynolds number. The arrow indicates increasing values of the Reynolds number. 
A spatial growth of the cross-stream kinetic energy is clearly visible in the wake of the flat-plate. This spatial growth is stronger and stronger 
when increasing the Reynolds number but also more and more localized. Indeed the cross-stream kinetic energy displays a peak, for instance at $x \sim 7$ 
for the largest Reynolds number $\textit{Re}=90$ shown in the Figure. Downstream of this station it strongly decreases. The spatial growth of this global mode is thus clearly limited to the near wake of the plate.\\

When increasing the Reynolds number above the first critical Reynolds number $\textit{Re}_{c1}$, other global modes may get unstable. 
The determination of those modes and their critical Reynolds number is a first necessary step towards a better understanding of the 
non-linear flow state. Indeed, if various global modes get unstable in a narrow range of Reynolds number, the weakly non-linear analysis 
proposed by Meliga \textsl{et al.} for two unstable modes might be performed to determine the bifurcation diagram. Such analysis is out of the scope 
of the present paper but the linear stability analysis has been further carried out to determine the modes getting unstable up to the Reynolds number $\textit{Re}=120$.\\
\begin{figure}
\begin{tabular}{ll}
(a) & (b) \\
\includegraphics[scale=.58]{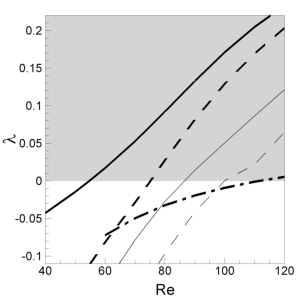} & \includegraphics[scale=.4]{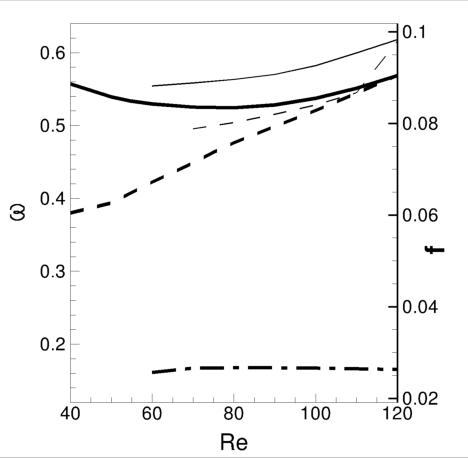}    
\end{tabular}
\vspace*{-0.0cm} 
\caption{Evolution of the stability properties with the Reynolds number. (a) Growth rate and (b) angular frequency of various global modes.
Solid lines: $(AS_y,S_z)$ modes; Dashed lines: $(S_y,S_z)$ modes; Dashed-dotted line : $(S_y,AS_z)$ modes. Parameter: $L=6$.}   
\label{fig:StabilityReynolds}
\end{figure}
The growth rate and frequency of those modes are depicted in Figure \ref{fig:StabilityReynolds}(a) and \ref{fig:StabilityReynolds}(b) respectively as a function of the Reynolds number. The thick solid lines in both figures correspond to the $(AS_y,S_z)$-symmetric mode so far described. When increasing the Reynolds number 
a second pair of complex eigenvalues, displayed by the thick dashed line in the figures, gets unstable for $\textit{Re}_{c2}=75$ as seen in Figure \ref{fig:StabilityReynolds}(a). As opposed to the first unstable mode, this second unstable mode is $(S_y,S_z)$-symmetric. 
Therefore this mode does not break any spatial symmetry of the base flow. The angular frequency at the critical Reynolds number 
is $\omega_{c2}=0.46$ and thus lower than the critical angular frequency $\omega_{c1}=0.53$. However when increasing the Reynolds number 
the angular frequency of this symmetry-preserving mode monotically increases and gets closer to the angular frequency of the $S_y$-breaking mode.
For the Reynolds number $\textit{Re}_{c3}=85$ a third pair of complex eigenvalues gets unstable, as shown by the thin solid line. This is the second $(AS_y,S_z)$-symmetric mode getting unstable. Its corresponding angular frequency is slightly larger than that of the first one, regardless of the value of the Reynolds number. 
The fourth destabilization of a pair of complex eigenvalues occurs at the Reynolds number $\textit{Re}_{c4}=100$ as shown by the dashed line. It 
corresponds again to a symmetry-preserving $(S_y,S_z)$ mode. Finally the fifth destabilitization of a pair of complex eigenvalues occurs for the Reynolds number $\textit{Re}_{c5}=115$ and corresponds to a $(S_y,AS_z)$-symmetric mode. This is the first unstable mode breaking the $S_z$-symmetry of the base flow. Its associated angular frequency, depicted by the thick dashed-dotted in Figure \ref{fig:StabilityReynolds}(b), 
is much lower than for the other unstable modes. The non-dimensional frequency
$f=0.041$ is lower than the frequency associated to global modes breaking the $S_y$s-ymmetry ($f=0.085$). Moreover it does not show any variation with respect to the Reynolds number, as if it was an inviscid mode. 

\begin{table}
\centering
\begin{tabular}{rcrcclr}
$i$ \vline & $\textit{Re}_{ci}$ & $\omega_{ci}$ \vline & $y$-symmetry & $z$-symmetry & \vline $\; f_{ci}$ & $\bar{f}_{ci}$ \\
\hline
$1$ \vline & $55$ & $0.53$ \vline & $AS$ & $S$ & \vline  $\; 0.084$ & $\; 0.504$ \\
$2$ \vline & $75$ & $0.46$ \vline & $S$ & $S$ & \vline   $\; 0.073$ & $\; 0.438$ \\
$3$ \vline & $86$ & $0.56$ \vline & $AS$ & $S$ & \vline  $\; 0.089$ & $\; 0.534$ \\
$4$ \vline & $100$ & $0.53$ \vline & $S$ & $S$ & \vline  $\; 0.084$ & $\; 0.504$ \\
$5$ \vline & $111$ & $0.16$ \vline & $S$ & $AS$ & \vline $\; 0.025$ & $\; \textbf{0.150}$
\end{tabular}
\caption{Summary of stability results for the flate-plate of aspect ratio $L=6$. Critical Reynolds number $\textit{Re}_{ci}$,
critical angular frequency $\omega_{ci}$ and spatial symmetries of the unstable global modes $i$. The non-dimensional frequencies 
$f_{ci}$ and $\bar{f}_{ci}= f_{ci} \, L$ are also indicated.}
\label{tab:stab}
\end{table}

\section{Flow stability around plate of smaller aspect ratio.}\label{sec:aspectratio}

The influence of the plate's aspect ratio on the flow stability is now investigated. The largest aspect ratio considered in the present paper is $L=6$ 
and has been studied in detail in the previous section. Attention is here paid on smaller aspect ratio's plates. The smallest aspect ratio considered 
in this study is $L=1$ and corresponds to a square plate. When the Reynolds number is varied in the range $50 \le \textit{Re} \le 150$, 
six global modes are identified as getting unstable. The neutral curves of those modes, i.e. the set of points in the control parameter's space $(L,\textit{Re})$ where their growth rate vanishes, have been determined.\\
\begin{figure}
\begin{tabular}{ll}
(a) & (b) \\
\includegraphics[scale=.26]{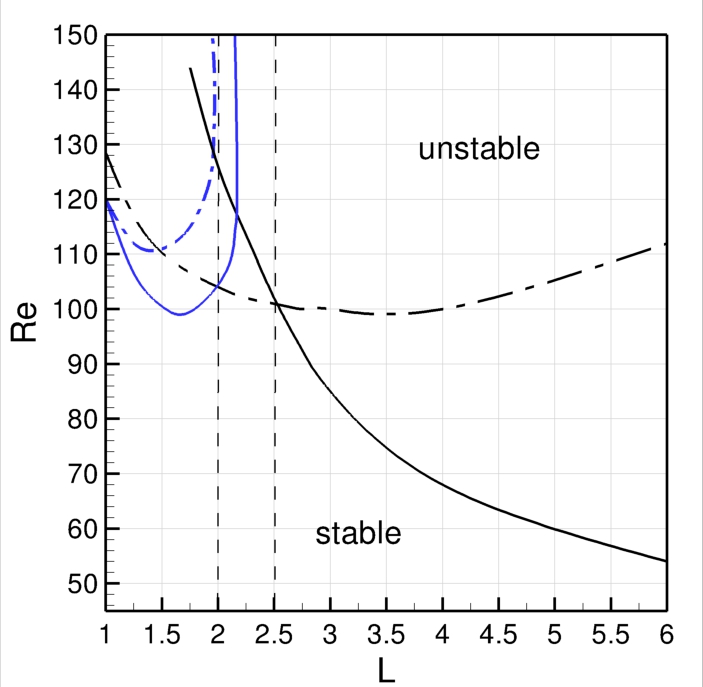} & \includegraphics[scale=.26]{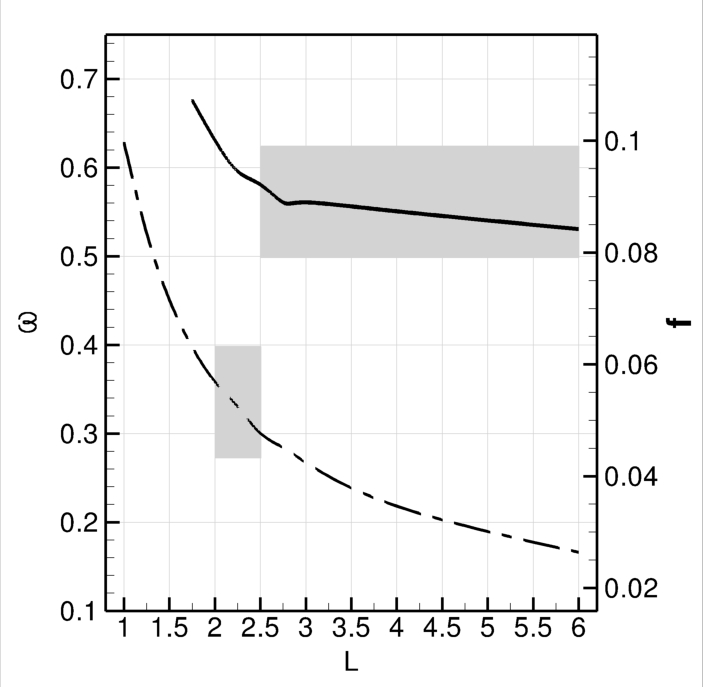} 
\end{tabular}
\vspace*{-0.0cm} \caption{(a) Neutral curves of four  global modes in the control parameter's space ($L$,$\textit{Re}$). 
The black lines stand for unsteady modes ($\omega \neq 0$) while the blue lines stand for  steady modes ($\omega = 0$).
Solid lines: ($AS_y,S_z$) modes; Dashed-dotted lines: ($S_y,AS_z$) modes. (b) Frequency of the marginal unsteady modes as a function of the aspect ratio }   
\label{fig:courbeneutresimple}       
\end{figure}
To ease the results discussion, only four neutral curves are first displayed in Figure \ref{fig:courbeneutresimple}(a). 
They correspond to global modes that gets first unstable, at a fixed aspect ratio, when increasing the Reynolds number. 
Blue curves in this figure are associated to steady modes ($\omega=0$) while black curves correspond to unsteady modes ($\omega \neq 0$). The angular frequency of the marginal unsteady modes is reported in Figure \ref{fig:courbeneutresimple}(b) as a function of the aspect ratio. 
Various destabilization scenarios are identified depending on which mode gets first unstable. They are now discussed, 
by starting with the largest values of the aspect ratio. \\

\begin{figure}
\begin{minipage}{1.0\textwidth}
\begin{tabular}{lr}
(a) & \\
& \includegraphics[scale=.5]{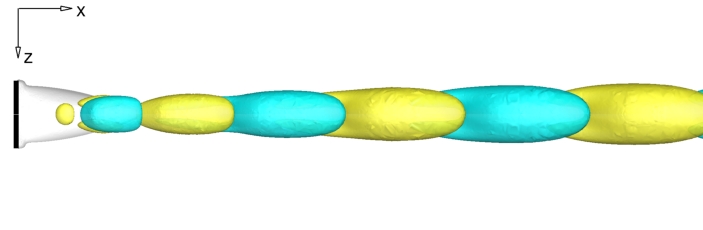} \\
(b) & \\
& \includegraphics[scale=.5]{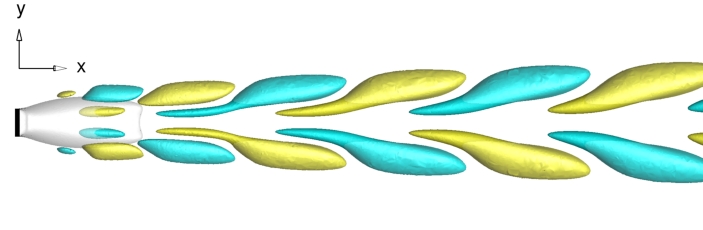} 
\end{tabular}
\end{minipage}
\vspace*{-0.0cm} \caption{Real part of the marginal unsteady global mode breaking the $S_y$-symmetry for $L=2.5$ and $\textit{Re}=101$. 
See the black solid line in Figure \ref{fig:courbeneutresimple}. Top (a) and side (b) views. Surfaces of positive (yellow) and negative (blue) 
streamwise isovelocity.}
\label{fig:ModeB}       
\end{figure}

For  $2.5 < L \le 6$ the first mode to get unstable when increasing the Reynolds number is unsteady and $(AS_y,S_z)$-symmetric. 
Its neutral curve is the black solid line in Figure \ref{fig:courbeneutresimple}(a). The steady wake flow is thus expected to bifurcate towards 
a time-periodic wake flow for which vortical structures are alternatively shed from the top and bottom shear layers. The spatial structure of the global mode 
displayed in Figure \ref{fig:ModeB} for the aspect ratio $L=2.5$ gives a first insight of the structures shed during a period of the phenomenom.
Note that it is quite similar to the one obtained for the aspect ratio $L=6$ and shown in Figure \ref{fig:Mode1}. 
The frequency of the marginal mode, depicted by the black solid line in Figure \ref{fig:courbeneutresimple}(b), varies only weakly with the aspect ratio, increasing from 
$f=0.084$ at $L=6$ to $f=0.092$ at $L=2.5$. This destabilization scenario is similar to the one obtained for two-dimensional bodies, except that the structures alternatively shed in the wake are three-dimensional.\\

Interestingly a decrease of the aspect ratio leads to a stabilization of the flow since the critical Reynolds number increases from $\textit{Re}_{c}= 55$ at $L=6$ to $\textit{Re}_{c}=101$ at $L=2.5$. Meanwhile, the critical Reynolds number of the $(S_y,AS_z)$-symmetric unsteady mode, depicted by the dashed-dotted line in Figures \ref{fig:courbeneutresimple}(a), slightly decreases. As a result two modes get simultaneously unstable for $(L,\textit{Re})=(2.5,101)$. One breaks the $S_y$-symmetry while the other one breaks the $S_z$-symmetry of the base flow. This is a codimension $2$ bifurcation point, called a Hopf-Hopf bifurcation since both modes are unsteady. 
In the vicinity of this point in the control parameter's space, the non-linear flow dynamics is expected to set in a slow manifold that 
can be unfolded around these two modes. The determination of the amplitude equations that govern the non-linear dynamics in the slow manifold 
is out of the scope of the present paper but could be done in the spirit of the weakly non-linear analysis proposed in \cite{Sipp2007},\cite{Meliga2009b}. Interestingly 
the frequency of the $S_z$-symmetry breaking mode is $f=0.048$, i.e. almost twice smaller 
than the frequency of the $S_y$-symmetry breaking mode $f=0.092$. This should lead to a strong 2:1 resonance and a particular form of the amplitude equations \cite{Meliga2012}. \\

\begin{figure}
\begin{minipage}{1.0\textwidth}
\begin{tabular}{lr}
(a) & \\
& \includegraphics[scale=.5]{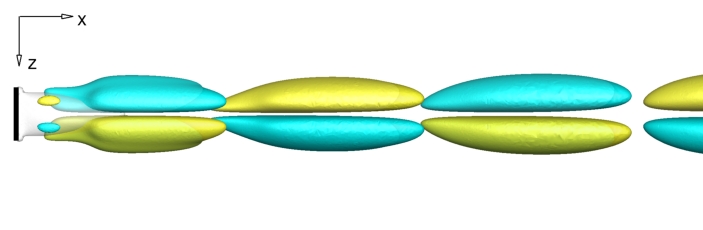} \\
(b) & \\
& \includegraphics[scale=.5]{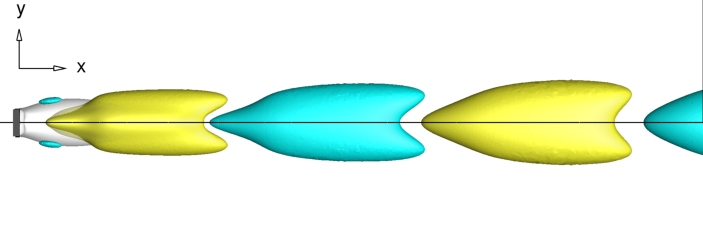} 
\end{tabular}
\end{minipage}
\vspace*{-0.0cm} \caption{Snapshot of the marginal unsteady global mode breaking the $S_z$-symmetry for $L=2$ and $\textit{Re}=105$. 
See the black dashed-dotted line in Figure \ref{fig:courbeneutre}.
Top (a), side (b) views.}
\label{fig:ModeC}       
\end{figure}

For $2 < L < 2.5$ the first mode to get unstable is now the unsteady $(S_y,AS_z)$-symmetric mode. The first bifurcation is therefore a Hopf bifurcation breaking 
the left/right symmetry of the base flow. The critical Reynolds number for this bifurcation weakly increases  from $\textit{Re}_{c} = 101$ at $L=2.5$ to $\textit{Re}_{c}=104$ at $L=2$. 
The frequency of this marginal mode, displayed in Figure \ref{fig:courbeneutresimple}(b) by the dashed-dotted line, increases from 
$f=0.048$ at $L=2.5$ to $f=0.057$ at $L=2$. The spatial structure of this mode is displayed in Figure \ref{fig:ModeC}. The side view shown in Figure \ref{fig:ModeC}(b) clearly highlights the top/bottom reflectional symmetry of the mode while the left/right anti-symmetry is visible in the top view displayed in Figure \ref{fig:ModeC}(a). Positive and negative velocity structures alternate in the streamwise direction. They are more elongated in the streamwise direction than for the $(AS_y,S_z)$-symmetric mode. \\

\begin{figure}
\begin{minipage}{1.0\textwidth}
\begin{tabular}{lr}
(a) & \\
& \includegraphics[scale=.5]{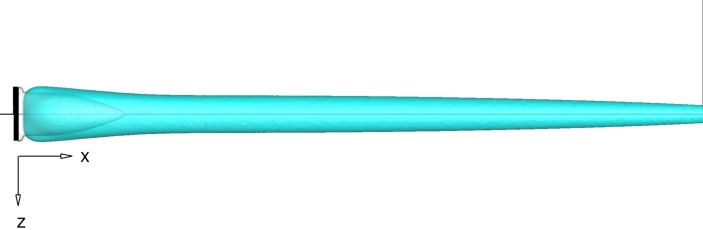} \\
(b) & \\
& \includegraphics[scale=.5]{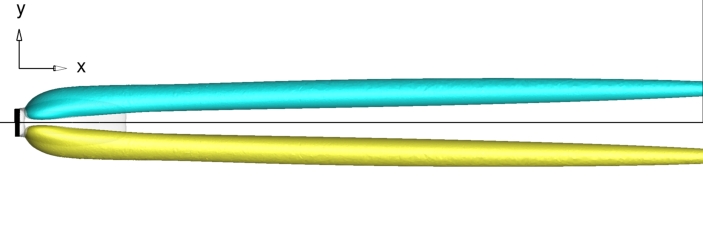} 
\end{tabular}
\end{minipage}
\vspace*{-0.0cm} \caption{Top (a) and side (b) views of the steady mode breaking the $S_y$-symmetry for $L=2$ and $\textit{Re}=105$. See the gray solid line in Figure \ref{fig:courbeneutresimple}.}
\label{fig:ModeA}       
\end{figure}

For $L=2$, two modes get simultaneously unstable, the unsteady $(S_y,AS_z)$-symmetric mode and a steady $(AS_y,S_z)$-symmetric mode whose neutral curve is the solid blue line.
This is a codimension $2$ bifurcation point called a pitchfork-Hopf bifurcation. The spatial structure of the steady mode is displayed in Figure \ref{fig:ModeA} by isosurfaces of the streamwise velocity. In the streamwise direction the structure is quite elongated and centered around the center axis line. The spatial pattern of this global mode
is quite similar to the first unstable global mode found in the wake of a sphere or disks \cite{Meliga2009a}. The effect of this mode is to deviate the wake out of the center axis line. In the present flow configuration a similar conclusion can be drawn except that the wake flow is deviated in the top or bottom direction. \\

For $1 < L < 2$ this steady $(AS_y,S_z)$-symmetric mode is the first one to get unstable. The first flow bifurcation is therefore a pitchfork bifurcation which 
breaks the top/bottom reflectional symmetry of the base flow. The critical Reynolds number of the mode breaking the $S_y$ symmetry 
first decreases to $\textit{Re}_{c} \sim 99$ at $L=1.7$ and then increases up to $\textit{Re}_{c}=120$ at $L=1$. Interestingly, the neutral curve (blue solid line) is almost vertical for $L>2$, indicating 
that this steady mode is always stable for large aspect ratio. One of the important results of the present paper is that steady modes breaking the reflectional symmetries get unstable only for low values of the aspect ratio. Note also that, if the bifurcation is supercritical, the steady symmetric wake flow is expected to bifurcate towards deviated wake flows above the marginal curve. 
The top and bottom deviated wake flows are equiprobable but 
small amplitude noise may force the flow to jump from one state to the other, leading to the bistability phenomenom. \\
Finally for $L=1$ two steady global modes get simultaneously unstable for the Reynolds number $\textit{Re}=120$. 
The second unstable steady mode has the symmetries $(S_y,AS_z)$ and its neutral curve is the dashed-dotted blue line. Starting from $L=1$ and increasing the aspect ratio, the present results show that steady mode breaking the \textit{major plane} symmetry get first unstable, as revealed by comparing the two blue curves in Figure \ref{fig:courbeneutresimple}(a).\\

\begin{figure}
\begin{tabular}{ll}
(a) & (b)  \\
\includegraphics[scale=.25]{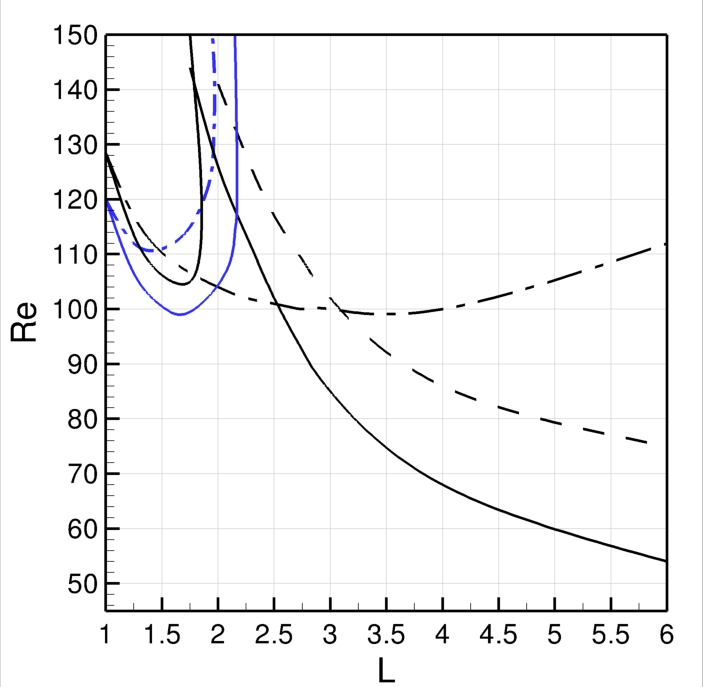} & \includegraphics[scale=.25]{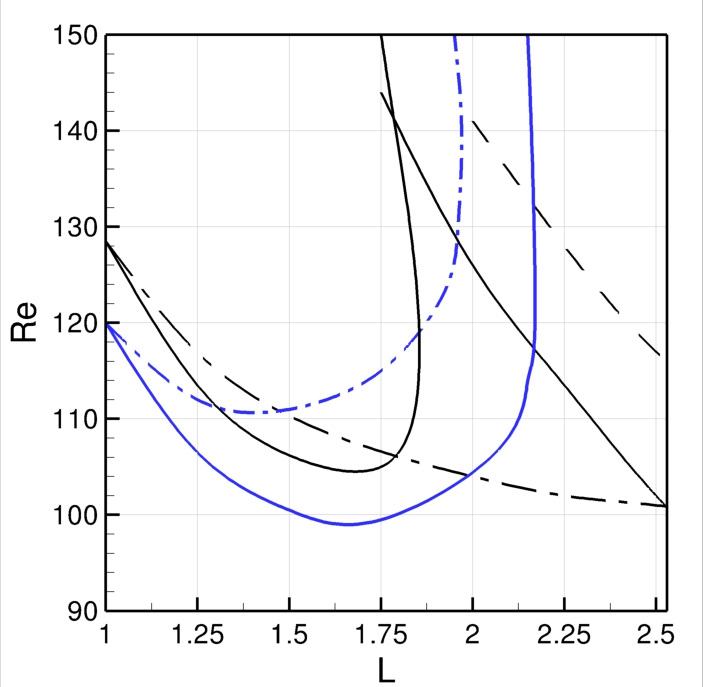} 
\end{tabular}
\vspace*{-0.0cm} \caption{Neutral curves of six global modes in the control parameter's space ($L$,$\textit{Re}$). (a) Large view (b) Close-up view.
The black lines stand for unsteady modes ($\omega \neq 0$) while the blue lines stand for steady modes ($\omega = 0$).
Solid lines: ($AS_y,S_z$) modes; Dashed lines: ($S_y,S_z$) modes; Dashed-dotted lines: ($S_y,AS_z$) modes.}   
\label{fig:courbesneutres}       
\end{figure}

\section{Conclusion}
The present results give a clear picture of the linear flow destabilization for plates of different aspect ratios. 
For small aspect ratio, $1 < L < 2$, the flow is destabilized by a steady mode breaking the top/bottom symmetry. 
For large aspect ratio, $L > 2.5$, the flow is destabilized by an unsteady mode also breaking the top/bottom symmetry. 
For intermediate aspect ratio,  $ 2 < L < 2.5$, the flow is also destabilized by an unsteady mode but breaking the left/right symmetry. \\
Such picture is expected to be more complex if non-linear effects 
as well as the existence of many unstable modes are taken into account. Linear stability results should thus be compared carefully with results of experiments or direct numerical simulations.
To give an insight into the complexity of the non-linear flow dynamics, the destabilization of other global modes has been tracked. 
Two additional neutral curves of global modes have been determined. These are the black dashed and dashed-dotted lines in Figure \ref{fig:courbesneutres} which correspond 
to ($S_y$,$S_z$)- and ($S_y$,$AS_z$)-symmetric modes respectively. \\
For large aspect ratio's plates the ($S_y$,$S_z$)-symmetric mode may play a role in the flow dynamics since this is the second mode getting unstable.
The corresponding critical Reynolds number (dashed line) increases when decreasing the aspect ratio, as for the neutral curve of the ($AS_y$,$S_z$)-symmetric mode (solid line).
For instance, for the aspect ratio $L=3$ three modes gets unstable in a narrow range of Reynolds number. The $(AS_y,S_z)$-symmetric mode gets first unstable around 
$\textit{Re}=85$, then ($S_y$,$S_z$)- and  ($S_y$,$AS_z$)-symmetric modes get simultaneously unstable around $\textit{Re}=100$. The coexistence of the three modes 
is expected to influence the non-linear flow dynamics. The number of modes getting unstable in a narrow range of values of the Reynolds number increases when considering small aspect ratio's plates, as seen in the close-up view displayed in Figure \ref{fig:courbesneutres}. 
Future works could be devoted to a better understanding of the nonlinear dynamics 
of wakes behind small aspect ratio plates, based on the knowledge of global modes getting unstable.





\section*{Bibliography}
\begin{thebibliography}{00}


\bibitem{Lawson2007}
Lawson N, Garry K, Faucompret N. An investigation of the flow characteristics in the bootdeck region of a scale model notchback saloon vehicle.{\it Proc. Inst. Mech. Eng., Part D (J. Automob. Eng.)} 2007;{\bf 221}(6):739-754.

\bibitem{Herry2011}
Herry B, Keirsbulck L, Paquet JB, Labrag L. Flow bistability downstream of three-dimensional double backward facing steps at zero-degree sideslip.{\it J. Fluids Eng.} 2011;{\bf 133}(5):054501.

\bibitem{Grandemange2013}
Grandemange M, Gohlke M, Cadot O. Bi-stability in the turbulent wake past parallelepiped bodies with various aspect ratios and wall effects. {\it Physics of Fluids}, 2013;{\bf 25}:095103.

\bibitem{Golubitsky1988}
Golubitsky M, Langford WF. Pattern formation and bistability in flow between counterrotating cylinders. {\it Physica D: Nonlinear Phenomena}, 1988;{\bf 32}:362-392.

\bibitem{Barkley2005}
Grandemange M, Gohlke M, Cadot O. Computational study of turbulent laminar patterns in Couette flow. {\it Phys. Rev. Lett.}, 2005;{\bf 94}:014502.

\bibitem{Williamson1996a}
Williamson CHK. Vortex dynamics in the cylinder wake. {\it Annual Review of Fluid Mechanics} 1996;{\bf 28}:477.

\bibitem{Zdrakovitch1997}
Zdrakovitch MM. Flow around circular cylinders. {\it Oxford University Press} 1997.

\bibitem{Thompson2001}
Thompson MC, Leweke T \& Provansal M. Kinematics and dynamics of sphere wake transition. {\it Journal of Fluids and Structures} 2001;{\bf 15}:575-585.

\bibitem{Fabre2008}
Fabre D, Auguste F, Magnaudet J. Bifurcations and symmetry breaking in the wake of axisymmetric bodies. {\it Physics of Fluids} 2008;{\bf 5}:051702.

\bibitem{Williamson1996b}
Williamson CHK. Three-dimensional wake transition. {\it Journal of Fluid Mechanics} 1996;{\bf 328}:345-407.

\bibitem{Barkley1996}
Barkley D, Henderson RD. Three-dimensional Floquet analysis of the wake of a circular cylinder. {\it Journal of Fluid Mechanics} 1996;{\bf 322}:215-241.

\bibitem{Thompson1996}
Thompson MC, Hourigan K \& Sheridan S. Three-dimensional instabilities in the wake of a circular cylinder. {\it Experimental Thermal and Fluid Science} 1996;{\bf 12}(2):190-196.

\bibitem{Grandemange2012}
Grandemange M, Cadot O, Gohlke M. Reflectional symmetry breaking of the separated flow over three-dimensional bluff bodies.{\it Physical review E} 2012;{\bf 86}:035302.


\bibitem{Schouveiler2001}
Schouveiler L, Provansal M. Periodic wakes of low aspect ratio cylinders 
with free hemispherical ends. {\it Journal of Fluids and Structures} 2001;{\bf 15}:565-573.

\bibitem{Provansal2004}
Provansal M, Schouveiler L, Leweke T. From the double vortex street behind a cylinder to the wake of a sphere. {\it European Journal of Mechanics B/Fluids} 2004;{\bf 23}:65-80.

\bibitem{Sheard2008}
Sheard GJ, Thompson MC \& Hourigan K. Flow normal to a short cylinder with hemispherical ends. {\it Physics of Fluids} 2008;{\bf 20}:041701.

\bibitem{Inoue2008}
Inoue O, Sakuragi A. Vortex shedding from a circular cylinder of finite length at low Reynolds numbers. {\it Physics of Fluids} 2008;{\bf 20}:033601.

\bibitem{Kuo1967}
Kuo YH, Baldwin LV. The formation of elliptic wakes. {\it Journal of Fluid Mechanics} 1967;{\bf 27}:353-360.

\bibitem{Kiya1999}
Kiya M, Abe Y. The formation of elliptic wakes. {\it Journal of Fluids and Structures} 1999;{\bf 13}:1041-1067.

\bibitem{Kiya2001}
Kiya M, Ishikawa H, Sakamoto H. Near-wake instabilities and vortex structures of three-dimensional bluff-bodies: a review. {\it Journal of Wind Engineering and Industrial Aerodynamics} 2001;{\bf 89}:1219-1232.

\bibitem{Ern2012}
Ern P, Risso F, Fabre D \& Magnaudet J. 
Wake-induced oscillatory paths of rising or falling rigid bodies.{\it Annual Review of Fluid Mechanics} 2012;{\bf 44}:97-121.


\bibitem{Jackson1987}
Jackson C P. A finite element study of the onset of vortex shedding in flow past various shaped bodies. {\it Journal of Fluid Mechanics} 1987;{\bf 182}:23-45.

\bibitem{Natarajan1993}
Natarajan R, Acrivos A. The instability of the steady flow past sphere and disks. {\it Journal of Fluid Mechanics} 1993;{\bf 254}:323-344.

\bibitem{Pier2008}
Pier B. Local and global instabilities in the wake of a sphere. {\it Journal of Fluid Mechanics} 2008;{\bf 603}:39-61.

\bibitem{Meliga2009a}
Meliga P, Chomaz JM, Sipp D. Unsteadiness in the wake of disks and spheres: Instability, receptivity and control using direct and adjoint global stability analysis. {\it Journal of Fluids and Structures} 2009;{\bf 25}:601-616.

\bibitem{Giannetti2007}
Giannetti F, Luchini P. Global stability of base and mean flows: a general approach and its applications to cylinder and open cavity flows. {\it Journal of Fluid Mechanics} 2007;{\bf 593}:333-358.

\bibitem{Sipp2007}
Sipp D, Lebedev A. Global stability of base and mean flows: a general approach and its applications to cylinder and open cavity flows. {\it Journal of Fluid Mechanics} 2007;{\bf 593}:333-358.

\bibitem{Marquet2008a}
Marquet O, Sipp D, Jacquin L. Sensitivty analysis and passive control of cylinder flow. {\it Journal of Fluid Mechanics} 2008;{\bf 615}:221-252.

\bibitem{Marquet2008b}
Marquet O, Sipp D, Jacquin L, Chomaz JM. Multiple timescale and sensitivity analysis for the passive control of cylinder flow. {\it $5^{th}$ AIAA Theoretical Fluid Mechanics Conference, Seattle, Washington} 23-26 June 2008.

\bibitem{Meliga2009b}
Meliga P, Chomaz JM, Sipp D. Global mode interaction and pattern selection in the wake of a disk: a weakly nonlinear expansion. {\it Journal of Fluid Mechanics} 2009;{\bf 633}:159.

\bibitem{Bagheri2009}
Bagheri S, Schlatter P, Schmid PJ, Henningson DS. Global stability of a jet in crossflow. {\it Journal of Fluid Mechanics} 2009;{\bf 624}:33-44.


\bibitem{Szaltys2012}
Szaltys P, Chrust M, Przadka A, Goujon-Durand S, Tuckerman L, Wesfreid JE. Nonlinear evolutions of instabilitibehind sphere and disks.{\it Journal of Fluid and Structures} 2012;{\bf 28}:483-487.

\bibitem{Ilak2012}
Ilak M, Schlatter P, Bagheri S, Henningson DS. Bifurcations and stability analysis of a jet in cross-flow: onset of global instability at a low velocity ratio.{\it Journal of Fluid Mechanics} 2012;{\bf 696}:94-121.um.tex'.


\bibitem{freefem}
Hecht F. New development in freefem++.{\it Journal of Numerical Mathematics} 2012;{\bf 20},3-4:251-265.

\bibitem{arpack}
Lehoucq RB, Sorensen DC, Yang C. ARPACK users guide: solution of large scale eigenvalue problems by implicitly restarted Arnoldi methods. 1997.

\bibitem{mumps}
Amestoy PR, Duff IS, L\'Excellent JY, Koster J. MUMPS: a general purpose distributed memory sparse solver.{\it Applied Parallel Computing. New Paradigms for HPC in Industry and Academia}, 2001:121-130. Springer Berlin Heidelberg.

\bibitem{tuckerman2000}
Tuckerman LS, Barkley D. Bifurcations analysis for timesteppers. 2000:453-466, Springer New York.

\bibitem{bagheri2009}
Bagheri S, Akervik E, Brandt L, Henningson DS. Matrix-free methods for the stability and control of boundary layers.{\it AIAA Journal} 2009,{\bf 47(5)}:1057-1068.

\bibitem{sipp2010}
Sipp D, Marquet O, Meliga P, Barbagallo A. Dynamics and control of global instabilities in open flows: a linearized approach.{\it App. Mech. Rev.} 2010,{\bf 63}:030801.

\bibitem{metis}
Karypis G, Kumar V. A fast and highly quality multilevel scheme for partioning irregular graphs.{\it SIAM Journal on Scientific Computing} 1999,{\bf 20}(1):359-392.

\bibitem{Meliga2012}
Meliga P, Gallaire F, Chomaz JM. A weakly nonlinear mechanism for mode selection in swirling jets. {\it Journal of Fluid Mechanics} 2012;{\bf 699}:216-262.

\end{thebibliography}


\section*{Acknowledgements}
This work was supported by the Agence Nationale de la Recherche under 
grants ANR-08-BLAN-0099 ENTOMOPTER and ANR-09-BLAN-0139 OBLIC. 
The one-year post-doctoral fellowship of the second author was funded under grant  
ANR-08-BLAN-0099 ENTOMOPTER. This work was granted access to the HPC resources
of CINES under the allocation 2011-c2011026675 made by GENCI.

\end{document}